\documentclass[conference]{IEEEtran}
\IEEEoverridecommandlockouts
\usepackage{cite}
\usepackage{amsmath,amssymb,amsfonts}
\usepackage{algorithmic}
\usepackage{graphicx}
\usepackage{textcomp}
\usepackage{xcolor}
\usepackage{makecell}
\def\BibTeX{{\rm B\kern-.05em{\sc i\kern-.025em b}\kern-.08em
    T\kern-.1667em\lower.7ex\hbox{E}\kern-.125emX}}
\begin{document}

\title{SoK: Machine Learning for Continuous Integration}   
% Automating Continuous Integration Tasks by Machine Learning 

\author{\IEEEauthorblockN{Ali Kazemi Arani}
\IEEEauthorblockA{\textit{School of Computer Science} \\
\textit{University of Adelaide}\\
Adelaide, Australia \\
ali.kazemiarani@adelaide.edu.au}
\and
\IEEEauthorblockN{Mansooreh Zahedi}
\IEEEauthorblockA{\textit{\makecell{School of Computing and\\Information Systems}} \\
\textit{University of Melbourne}\\
Melbourne, Australia \\
mansooreh.zahedi@unimelb.edu.au}
\and
\IEEEauthorblockN{Triet Huynh Minh Le}
\IEEEauthorblockA{\textit{School of Computer Science} \\
\textit{University of Adelaide}\\
Adelaide, Australia \\
triet.h.le@adelaide.edu.au}
\and
\IEEEauthorblockN{Muhammad Ali Babar}
\IEEEauthorblockA{\textit{School of Computer Science} \\
\textit{University of Adelaide}\\
Adelaide, Australia \\
ali.babar@adelaide.edu.au}
}

\maketitle

\begin{abstract}
Continuous Integration (CI) has become a well-established software development practice for automatically and continuously integrating code changes during software development. An increasing number of Machine Learning (ML) based approaches for automation of CI phases are being reported in the literature. It is timely and relevant to provide a Systemization of Knowledge (SoK) of ML-based approaches for CI phases. This paper reports an SoK of different aspects of the use of ML for CI. Our systematic analysis also highlights the deficiencies of the existing ML-based solutions that can be improved for advancing the state-of-the-art.  
  
%has been increasingly used and has also shown promising results in harnessing large-scale data in CI environments to automate various CI phases.
% With the increased agility provided by the CI process, and the recent advances in DevOps for providing fast feedback to the developers, the use of Machine Learning (ML) algorithms in the CI environment has seen a significant increase in popularity.
%As the use of ML for CI becomes mainstream, a systematic review of the techniques employed in this emerging area is necessary. This study %presents a Systematization of Knowledge (SoK) of the CI phases that have been automated by ML in the literature and the practices followed by the state-of-the-art ML-based solutions for each identified phase.
%We also discuss the deficiency in the current solutions, providing insights for advancing this field.

\end{abstract}


\section{Introduction}
% DevOps is a software development methodology that makes a stronger connection between Development (Dev) and Operation (Ops) teams. The lack of connection between Dev and Ops teams has been identified as a significant issue by large software companies, as reported in literature \cite{debois2011devops}. In other words, DevOps integrates the development and deployment processes of software products into production environments \cite{fitzgerald2017continuous}. DevOps also offers additional advantages such as the ability to rapidly and continuously respond to changing user demands and requirements. This is achieved through the use of automation in the build, testing, and deployment phases of the software development process. Additionally, DevOps enables rapid feedback from end-users to be returned to the development teams, allowing for quick adjustments and improvements to be made to the software \cite{figalist2019supporting}. 

\par In recent years, the software development industry has seen a significant shift towards the adoption of Continuous Integration (CI) practices. The CI process is a software development approach that aims to improve the speed and reliability of software delivery by continuously integrating code changes into a shared repository. The goal of the CI process is to catch and fix issues and bugs early in the development process before they become major problems that are harder and more expensive to resolve \cite{shahin2017continuous}.
\par The growing popularity of CI and DevOps, along with the increasing volume and complexity of data involved in these processes, have motivated researchers to propose Machine Learning (ML) based solutions for automating CI phases \cite{shahin2017continuous} and taking one more step toward enabling the AIOps~\cite{sen2021devops}.
% The large scale and high dimension of data in CI make it an attractive domain for ML-based automation \cite{shahin2017continuous}.
% CI paradigm allows teams/organisations to continuously deliver high-quality software in a fast and reliable manner. 
ML methods can support the fast feedback loop in CI by analysing development data, deployment log files, and data from the operating environment and making automated decisions which are the exclusive benefits of ML methods \cite{figalist2019supporting}. 
These techniques can also be used for efficiently predicting the outcome of complex tasks. For example, ML methods can predict different types of software defects based on previous versions of code without running the current version \cite{kawalerowicz2021continuous}. Furthermore, ML-based solutions can provide accurate estimations, facilitate adapting to frequent changes \cite{pospieszny2017software}, and assist software engineers in timely decision-making processes \cite{figalist2019supporting}. The benefits highlighted for ML-based solutions in CI environments would facilitate a reduction in human intervention and enhance the efficiency of CI services in cloud-based environments, which are essential requirements for cloud-based environments. \cite{rimal2011architectural}. 
% Based on the numerous benefits that presented above, we decided to focus solely on ML-based methods in this study. 
\par Given the variety of employed techniques in applying ML solutions in CI, and growing interest in this domain, it is necessary to systematically identify state-of-the-art practices used for automating CI tasks through ML methods. This body of knowledge can provide a valuable reference for practitioners and researchers to understand the potential of available ML techniques and make an informed decision to apply ML models in real-world CI environments.
To the best of our knowledge, there is no existing Systemization of Knowledge (SoK) of the state-of-the-art ML practices for CI.

% \par In our study, we focused on comprehensively investigating the development phases of the state-of-the-art ML-based solutions for each CI phase. Our approach includes an analysis of raw data by synthesizing them. Our work is distinguished from other studies in that we extract data based on considering both the training phases of ML models and the CI phase of CSE practices. This approach allows for a more comprehensive understanding of the application of ML in CI and the specific techniques used in the field. Our study aims to provide valuable insights for researchers and practitioners in the field and to identify areas where further research is needed to improve the effectiveness of ML-based solutions in CI.
\par To bridge this gap, we review existing ML solutions that have been developed for CI phases in the last decade.
Specifically, we analyze relevant scientific articles on the topic published in peer-reviewed venues.
We first identify the CI phases that have been automated by ML.
% Secondly, we present the published studies that automated CI tasks as the state-of-the-art ML approach in each CI phase.
Moreover, we provide insights into key ML practices such as data/feature engineering, learning algorithms, and evaluation procedures, employed in these state-of-the-art ML solutions.
Such knowledge can serve as a guide for practitioners to adopt/develop high-performing ML solutions to automate CI phases at scale.
We also discuss areas for improving the utilization of the current ML solutions in real-world CI settings.

 \section{Background}
 \begin{figure}[bt]
\centerline{\includegraphics[width=0.9\columnwidth]{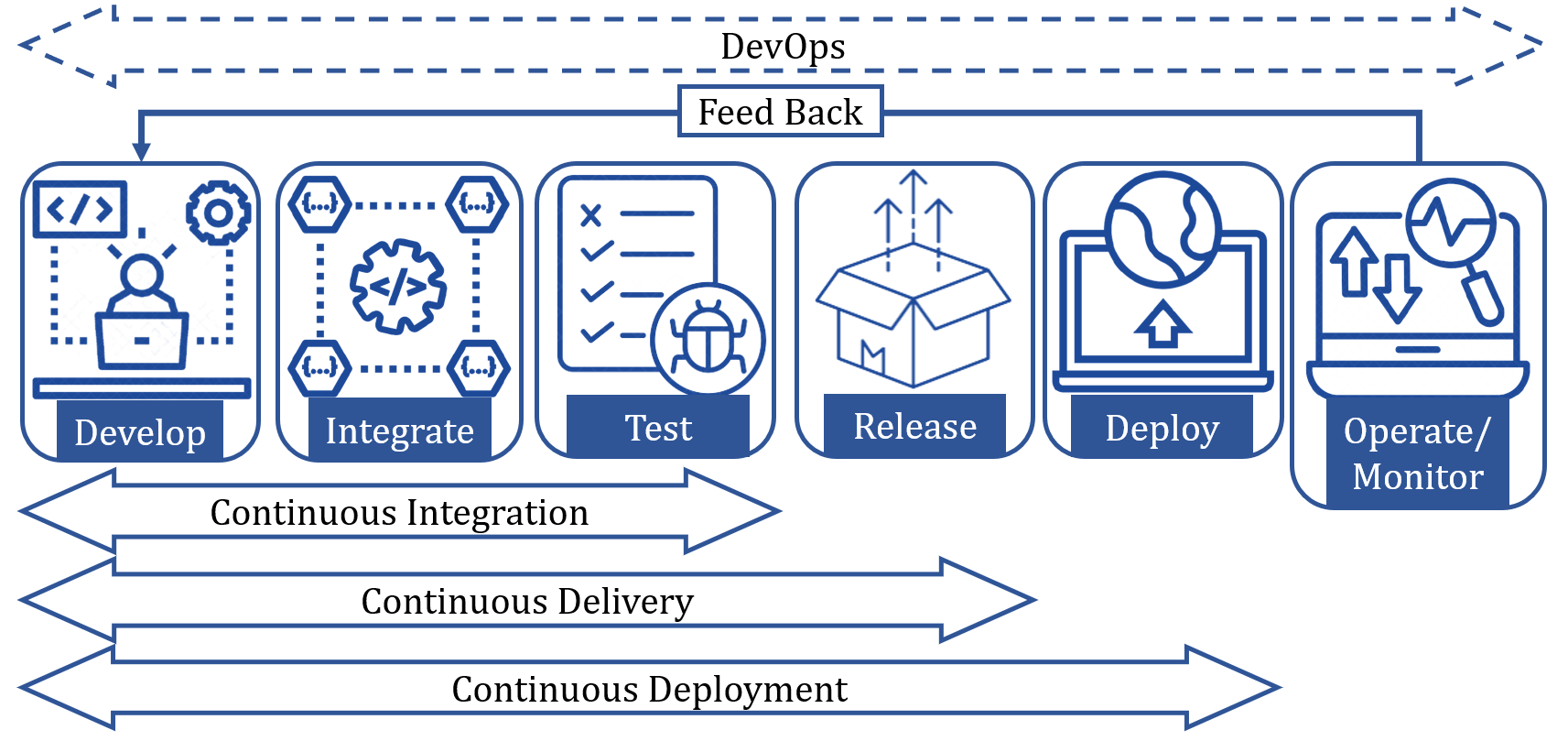}}
\vspace{-2mm}
\caption{The key steps and concepts of Continuous Software Engineering.
    \textbf{Notes:} \emph{Rectangles:} Steps. \emph{Two-way arrows:} Automatic phases. \emph{Dashed arrows:} Spanning multiple steps.}
\label{Fig:CSE_DevOps}
\vspace{-5mm}
\end{figure}

\par This review focuses on CI, which is a part of Continuous Software Engineering (CSE) that aims to provide quick and frequent feedback from customers' experiences, as well as software operations and maintenance \cite{fitzgerald2017continuous}. According to Figure \ref{Fig:CSE_DevOps}, CI is the first step of CSE that comes before the Continuous DElivery (CDE) and Continuous Deployment (CD) phases. Therefore, improving the performance of CI will positively affect the whole process of CSE.
CI includes software building, testing, and validation \cite{fitzgerald2017continuous}. 
% Members of a development team build software by integrating and merging their source code frequently (e.g., multiple times in a day) \cite{shahin2017continuous}. The frequent automatic tests and quick feedback for developers are crucial to prevent the propagation of issues to the delivery phase or affecting the development process by other developers \cite{fitzgerald2017continuous}.
This process involves multiple integration within a day and requires frequent automatic testing and prompt feedback to prevent issues from propagating to the delivery phase or affecting the development process of other team members \cite{shahin2017continuous}.
Lastly, developers receive feedback on detected bugs or performance issues through the validation phase in CI and also by maintaining (monitoring in operations) deployed software \cite{debois2011devops}.

\par Most of the existing systematic review/mapping studies on the application of ML methods in CI have mainly focused on the Test Case Prioritization (TCP) and Test Case Selection (TCS) methods.
Pan et al. \cite{pan2022test} investigated the employed feature sets and evaluation metrics in the presented ML-based solution for TCP and TCS in CI environments. Also, in \cite{lima2020test}, Lima et al. only focused on reviewing TCP methods in CI and reported the types of ML approaches and the evaluation methods for the task. In contrast, our review did not limit the search to a specific solution in CI and instead investigated all ML-based methods in CI environments. This approach allows for a more comprehensive understanding of the application of ML in CI and the techniques employed in the state-of-the-art methods in the field. The aim of this study is to provide valuable insights for researchers and practitioners in the field and to identify areas where further research is needed to improve the effectiveness of ML-based solutions in CI.

\section{Methodology}
\label{Scopus}
% \par For collecting the relevant studies in the area of application of ML-based methods in CI, we followed the guidelines for conducting a Systematic Literature Review (SLR) provided in \cite{kitchenham2007guidelines}. This includes a rigorous and systematic search for relevant studies in the field, along with an extensive screening and selection process to ensure that only the most relevant and high-quality studies are included in our review. By following these guidelines, we aim to ensure that our study is based on a comprehensive and representative sample of the literature in the field and that the results of our study are reliable and valid.
\par We aim to provide a Systematization of Knowledge of the application of ML-based methods in CI by following guidelines for Systematic Literature Review (SLR)~\cite{kitchenham2007guidelines}. We performed a rigorous and systematic search for relevant studies and used various criteria for ensuring the quality of the studies included in our review.
To achieve our aim, we investigated the two following Research Questions (RQs).
% We answered the two following questions to identify the key ML applications in CI and then provide insights into the techniques employed for developing state-of-the-art ML solutions in this area.

% In this study, we aim to investigate the employed ML-based techniques in the latest published (state-of-the-art) works in each automated CI task. Therefore, we answer two questions to first identify the CI phases and automated tasks and second extract details on the employed techniques in developing ML models:

\begin{itemize}
    \item \textbf{RQ1}: What are the CI phases that have been addressed by ML-based solutions?
    \item \textbf{RQ2}: What are the techniques employed in the development of state-of-the-art ML solutions for each CI phase?
\end{itemize}

The answers to these questions would help highlight the CI phases for which ML has shown potential and the current best practices for leveraging ML for automating CI phases.
% The answers to these questions are expected to provide a comprehensive overview of the current state of research on the application of ML in CI and to identify the key ML methods used in the field.
\subsection{Study selection}
% For finding the relevant studies we used the Scopus\footnote{https://www.scopus.com/} indexing system by running a search string. We selected the Scopus database because a large number of journals and conferences in the Software Engineering and Computer Science domains that typically publish CI-related studies are indexed in this database. This is supported by literature \cite{kitchenham2010systematic, daneva2014empirical} which highlights the coverage of Scopus as a comprehensive source of scientific and technical papers in the Computer Science and Engineering domains.
We used Scopus\footnote{https://www.scopus.com/} as the main source for finding relevant studies because it includes many journals and conferences in Software Engineering and Computer Science that are relevant to CI-related studies~\cite{kitchenham2010systematic, daneva2014empirical}.
% highlighting its coverage as a comprehensive source for papers in Computer Science and Engineering domains.

We then designed a search string to retrieve relevant studies on Scopus. Our search string was designed based on three segments, including \textbf{A)} Machine Learning, Artificial Intelligence and associated synonyms, \textbf{B)} the synonyms for CI, DevOps and CSE and \textbf{C)} the synonyms for “software”, “information systems”, “information technology”, “cloud”, and “service engineering” terms. This approach ensured that our search was comprehensive and covered a wide range of relevant studies in the field. By combining these three segments in our search string helped us capture most of the relevant studies on the application of ML-based methods in CI and DevOps.

%%%%%%%%%%%%%%%%%%%%%%%%%%%%%%%%%%%%%%
%%                                  %%
%%      Inclusion and Exclusion     %%
%%                                  %%
%%%%%%%%%%%%%%%%%%%%%%%%%%%%%%%%%%%%%%

\subsection{Inclusion and Exclusion Criteria}
\label{Sec:InclusionExclusion}
For identifying and removing irrelevant and low-quality studies, we defined the following inclusion criteria and selected papers based on these criteria.

\begin{enumerate}
    \item Research papers that were longer than four pages
    \item The full text of the papers was available in English
    \item The key topic of the papers was the application of ML-based methods in CI
\end{enumerate}
% By using these inclusion and exclusion criteria, we aimed to ensure that only high-quality and relevant studies are included in our review and that the results of our study are based on a comprehensive and representative sample of the literature in the field.

Using the explained settings, we ran the search string on the Scopus indexing system and retrieved 1,662 studies. After applying the inclusion criteria and conducting snowballing methods for reducing the risk of missing studies \cite{wohlin2014guidelines}, we obtained a final list of 44 related studies.
Intuitively, the latest studies usually present state-of-the-art solutions.
However, some of the latest studies might not include state-of-the-art ML solutions.
% To ensure the relevance of the selected studies for RQ2, we read the full-text and ensured that each paper presented a novel approach and showed that the proposed approach was better than the existing state-of-the-art one in the respective literature.
To ensure relevance for RQ2, we thoroughly examined the full text of each paper to find each paper presented a novel approach and showed the superiority of their method over current state-of-the-art in the relevant literature.
This would assist practitioners and researchers in determining the appropriate contexts for utilizing their ML-based methods in automating CI tasks.
% Secondly, we select the state-of-the-art studies by ranking the studies based on the rank of their publishers in the Core\footnote{https://www.core.edu.au/} system and in each CI task.

%%%%%%%%%%%%%%%%%%%%%%%%%%%%%%%%%%%%%%
%%                                  %%
%%  Data extraction and synthesis   %%
%%                                  %%
%%%%%%%%%%%%%%%%%%%%%%%%%%%%%%%%%%%%%%

\subsection{Data extraction and data synthesis}
\label{Sec:Synth}
% Based on the SLR guidelines, we defined a data extraction form for systematically extracting relevant information from the list of selected studies to answer the research questions (RQs) \cite{kitchenham2007guidelines}. For finding the answer of first research question we synthesized and analyzed the extracted data based on thematic analysis \cite{braun2006using} guidelines. This approach allowed us to systematically extract and organize the data from the selected studies and to identify patterns and themes in the data. Also, for answering the second question we followed the statistical analysis.
Based on SLR guidelines, we created a data extraction form to systematically extract information from selected studies to answer the RQs~\cite{kitchenham2007guidelines}. We used thematic analysis \cite{braun2006using} to synthesize and analyze data for RQ1 and statistical analysis for RQ2. This approach allowed us to extract and organize data from selected studies, identify patterns and themes, and answer the RQs.

\section{Results}
\subsection{RQ1: CI phases and ML applications}
\par
\begin{figure}[bt]
\centering
    \includegraphics[width=0.8\columnwidth,keepaspectratio]{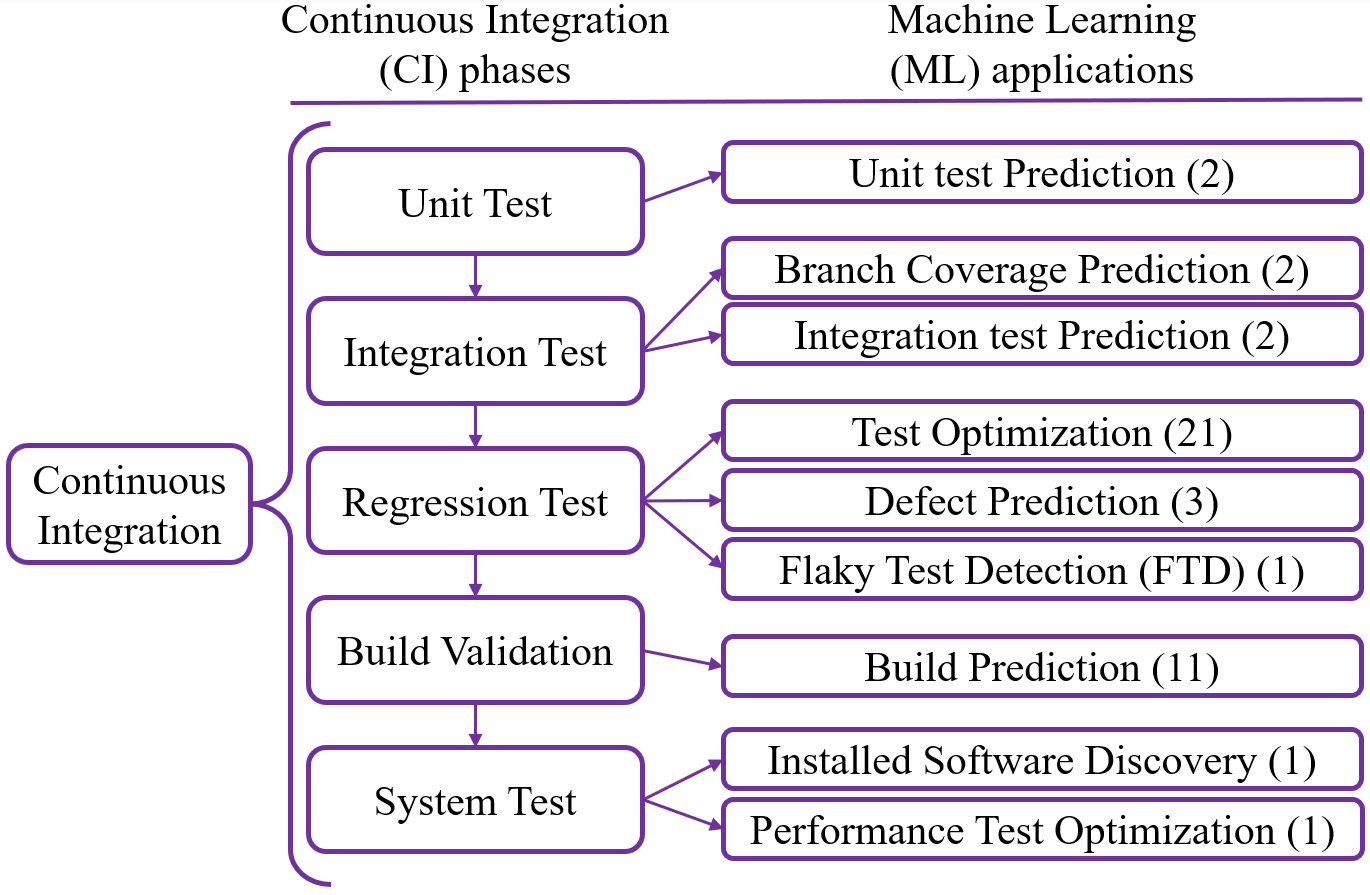}
\vspace{-4mm}
\caption{Relation between the identified CI phases and application of ML-based methods. \textbf{Note:} Numbers of published papers are presented in parenthesis.}

\label{Fig:CITasks}
\vspace{-5mm}
\end{figure}

 Based on reviewing and analyzing the selected papers, we identified five CI testing phases automated by ML.
Figure \ref{Fig:CITasks} presents the application of ML-based methods to the five CI phases.
% The number of published papers in each CI phase is presented in parentheses in Figure \ref{Fig:CITasks}. 
The descriptions of these five identified phases based on their sequence in the CI pipeline are presented below.

% \par \textbf{Unit Test (UT):}
\par \subsubsection{Unit Test (UT)}
UT is the first step in the chain of events in the CI pipeline \cite{stolberg2009enabling}. This test checks and validates the developed code in an isolated environment when a developer check-in (commits) new or modified compiled code and builds it \cite{zhou2012depth}. To keep the master branch free from error, developers commit all the changes on a developers' branch at this step \cite{martins2021supervised}. Two out of the 44 selected studies presented ML-based solutions to reduce the risk of buggy software releases by predicting the outcome of the unit tests without explicitly running them. Lee et al. \cite{lee2019classifying}, proposed a method for predicting the outcomes of unit tests using ML, and demonstrate its effectiveness in estimating the state of the alarms raised by static checkers. Vig et al. \cite{vig2018test} proposed a method for estimating the required test efforts by predicting the outcome of unit tests using ML. They compared different ML methods and feature sets to evaluate the effectiveness of their method.

% \par \textbf{Integration Test (IT):}
\par \subsubsection{Integration Test (IT)}
IT is a CI step where the newly developed modules are integrated with the existing ones on the same branch after they have been validated through unit testing \cite{abdalkareem2020machine}. The purpose of IT is to verify that all the modules of a system or subsystem work together seamlessly. In this phase, the utilization of ML techniques can aid in forecasting the outcomes of test cases \cite{abdalkareem2020machine}, and identifying the branch coverage of test cases to detect bugs with fewer test runs without missing any changes that need to be tested \cite{grano2019branch}.
% \par \textbf{Regression Test (RT):}
\par \subsubsection{Regression Test (RT)}
After validating and testing the integration of new or modified software units with other software units, the current version needs to be tested entirely based on the previously designed test cases \cite{ali2019enhanced}. 
% The tests in regression testing encompass various types of testing such as structural and functional testing \cite{elsner2021empirically}. 
The goal of all 25 RT studies can be classified into ``Test Optimization'' studies by employing Test Case Prioritization (TCP) and Test Case Selection (TCS) strategies, predicting the outcome of the test suites for committed codes as
``Defect Prediction'' studies, and ``Detecting Flaky Tests''. Flaky tests are the tests that produce an inconsistent pass or fail results and make the CI pipeline unreliable \cite{parry2022evaluating}. In the field of TCP and TCS studies, the objective of ML models is to prioritize the test cases or select a subset of them that are more likely to uncover bugs and defects in the software under development \cite{yaraghi2022scalable}. This is achieved by analyzing previous changes in source code and the results of previous test runs to identify patterns that indicate a higher likelihood of uncovering defects.

% \par \textbf{Build Validation (BV):}
\par \subsubsection{Build Validation (BV)}
According to the outcome of the previous CI phases, developers can be relatively sure about the functionalities and performance of developed software units. So, this version of the software is eligible for merging into the master branch. 
%as the final product is built based on the existing developed units on this branch.
The BV test is to ensure the stability of integrated code before releasing the software and handing it over to a system testing team \cite{finlay2014data}. Due to the high computational cost of building software products \cite{xie2018cutting}, predicting the build outcome is the main target of the presented ML-based solutions in the BV phase. Build prediction aims to reduce the resource usage and building time of software by predicting the outcome of the build process. The predictions are made based on analyzing within-project data or cross-project data \cite{xia2017empirical}. Nine out of 11 studies in this CI phase used batch data of the CI pipeline to train their models. However, the authors of \cite{finlay2014data} and \cite{hassan2017change} trained their ML models using a stream of data and changes in data, respectively.

% \par \textbf{System Test (ST):}
\subsubsection{System Test (ST)}
In this step, all CI-related and quality aspects including the performance, functionalities, and compatibility of different software units are tested. This phase can be performed by a portion of the users or an internal system testing team \cite{vemulapati2019ai}. For example, beta testers (users who receive the developed software before official releases) are asked to provide their experience using the the fully integrated software product \cite{porres2020automatic}. ML-based solutions in this phase target the discovery of installed software for ensuring compliance, security, and efficiency of the system under test \cite{byrne2020praxi}, and the optimization of test suites for detecting performance defects in developed software products \cite{porres2020automatic}.

\begin{table*}[tb]
\caption{Employed techniques, data and characteristics of the state-of-the-art solutions. 
}
\vspace{-2mm}
\label{table:MLMethods}
\begin{center}
\vspace{-5mm}
\resizebox{\textwidth}{!}{%
\begin{tabular}{|l|c|c|c|c|c|c|}

% {|p{1.7cm}|p{2cm}|p{2.2cm}|p{2.5cm}|p{2cm}|p{1.65cm}|p{3cm}|}
\hline
\textbf{Reference} & \makecell{\textbf{CI phase} - \\\textbf{ML application}}  & \makecell{\textbf{Data source} - \\\textbf{Data Type}}      & \makecell{\textbf{Data} \\\textbf{Preparation}}    & \makecell{\textbf{Feature} \\\textbf{Engineering}} & \makecell{\textbf{Learning} \\\textbf{algorithm(s)}}    & \makecell{\textbf{Evaluation methods}\\ \textbf{- Metrics}}                   \\ \hline
    \textbf{Lee et al. \cite{lee2019classifying}} & \makecell{UT - \\Test Prediction} & \makecell{Samsung projects (Industrial) - \\ Source code} & \makecell{Filtering - \\selecting source code \\lines} & \makecell{Tagging - \\word2vec} & \makecell{NN\\ (Supervised)} & \makecell{Random K-fold - \\F1, Precision, \\Recall and Accuracy} \\\hline
 \textbf{Grano et al. \cite{grano2019branch}} & \makecell{IT - \\Branch Coverage \\Prediction} & \makecell{Google and Apache (OS) - \\Source code and Code MD} & \makecell{Not reported} & \makecell{Scaling - \\Normalization} & \makecell{RF, NN, SVM \\(Supervised)} & \makecell{Random K-fold - \\Mean Absolute Error \\(MAE)} \\\hline
 \textbf{Abdalkareem et al.\cite{abdalkareem2020machine}} & \makecell{IT - \\Test Prediction} & \makecell{GitHub (OS) - \\Commit MD} & \makecell{Balancing - \\Weka} & \makecell{Weighting - \\TF/IDF}  & \makecell{DT (C4.5) \\(Supervised)} & \makecell{Random K-fold - \\F1, Precision, \\Recall and AUC} \\\hline
 
 \textbf{Yaraghi et al. \cite{yaraghi2022scalable}} & \makecell{RT - \\Test Optimization} & \makecell{GitHub (OS) - \\Build logs, Test code and \\Source code MD} & \makecell{Building data - \\Dependency graph of\\ source code files\\ and test cases} & \makecell{Scaling \\and Weighting - \\Normalization \\and TF/IDF} & \makecell{RF, SVM, XGBoost \\(Supervised)} & \makecell{Random K-fold - \\APFD\textsubscript{C}} \\\hline

 \textbf{Pan et al.\cite{pan2021continuous}} & \makecell{RT - \\Defect Prediction} & \makecell{GitHub (OS) - \\Code MD and Test MD} & \makecell{Filtering - \\Based on last two outcome \\of the test cases} & \makecell{Not reported} & \makecell{DT, RF, MLP, \\NB, SVM, LR \\(Supervised)} & \makecell{Sorted K-fold - \\F1, AUC, MCC\\ and G-Measure} \\\hline
 \textbf{Parry et al.\cite{parry2022evaluating}} & \makecell{RT - \\Flaky Test Detection} & \makecell{Apache (OS) - \\Test MD and Test Code} & \makecell{Balancing - \\Tomek, SMOTE, \\Edited nearest neighbors} & \makecell{Scaling - \\Standardization} & \makecell{RF \\(Supervised)} & \makecell{Sorted K-fold - \\F1, Precision, Recall, \\FP, FN, and TP} \\\hline
 \textbf{Saidani et al. \cite{saidani2022improving}} & \makecell{BV - \\Build Prediction} & \makecell{TravisTorrent (OS) - \\Source code MD and Test MD} & \makecell{Balancing - \\SMOTE} & \makecell{Not reported} & \makecell{LSTM, RNN \\(Supervised)} & \makecell{Sorted K-fold - \\F1, Precision, Recall, \\Accuracy and AUC} \\\hline
 \textbf{Byrne et al. \cite{byrne2020praxi}} & \makecell{ST - \\Installed Software \\Discovery} & \makecell{Simulated cloud environment - \\System logs} & \makecell{Filtering - \\Noise reduction} & \makecell{Tagging - \\Columbus} & \makecell{RL (Vowpal Wabbit) \\(Supervised)} & \makecell{Random K-fold - \\F1, Precision\\ and Recall} \\\hline
 \textbf{Porres et al.\cite{porres2020automatic}} & \makecell{ST - \\Performance Test \\Optimization} & \makecell{Simulated testing scenarios - \\Test MD} & \makecell{Not reported} & \makecell{Scaling - \\Normalization}  & \makecell{DNN \\(Supervised)} & \makecell{Gradually evaluation - \\Positive predictive value \\(PPV)} \\ \hline
 \multicolumn{7}{l}
{\makecell[l]{\textbf{Notes:} OS = Open Source, MD = Meta Data, NN = Neural Network, DT = Decision Tree, RF = Random Forest, DNN = Deep Neural Network, RNN = Recursive Neural Network, \\LR = Linear Regression, RL = Reinforcement Learning.}}
\end{tabular}%
}
\vspace{-8mm}
\end{center}
\end{table*}

\subsection{RQ2: State-of-the-art ML-based methods for automating the CI testing phases}
% In order to identify the state-of-the-art ML-based methods in the field of CI testing, we compiled a list of papers for each phase and each ML application and selected the most recent paper published in peer-reviewed venues.
According to the results of the RQ1, we identified the state-of-the-art ML solutions in each CI automated task. The list of the selected papers, properties of the collected data, as well as the employed techniques for training and evaluating the ML methods are analysed and presented in Table \ref{table:MLMethods}.
The key lessons learned in different phases of developing state-of-the-art ML solutions are presented hereafter.

% \par \textbf{Data source and data preparation:}
\par \subsubsection{Data source and data preparation}
According to the reviewed studies, a majority of the papers in the field of CI testing (24 out of 44), employed Open-Source (OS) projects.
% Table \ref{table:MLMethods}, presents the popularity of the OS projects in the state-of-the-art methods as well.
The popularity of the OS projects is due to the availability of different types of OS project data, including source code, test code, and their metadata, as well as access to the data of cloud-based tools such as Travis CI
% \footnote{https://www.travis-ci.com/}
and GitHub. For example, Pan et al. \cite{pan2021continuous}, evaluated their model on 242 OS projects, and Parry et al. \cite{parry2022evaluating} assessed their model on 26 open-source Python projects. 
% \par Also, collecting the same data based on reviewing previous studies is possible in publicly available OS projects. For instance, \cite{yaraghi2022scalable} collected their training data based on three published works in the area of TCP.

Additionally, Table \ref{table:MLMethods} shows different data preparation techniques used to ensure the quality of the input data for the ML models.
% The first data preparation approach is data balancing using available re-sampling algorithms, such as re-sampling methods in Weka \cite{abdalkareem2020machine}, oversampling to increase the number of instances in the minority class (e.g., SMOTE) \cite{chawla2002smote}, or by using the under-sampling technique (e.g., TOMEK) \cite{dalla2021within}. Researchers can also benefit from the combination of both over- and under-sampling methods by using the SMOGN technique introduced in \cite{sharif2021deeporder}.
The first step in data preparation for ML is to balance the data using various re-sampling algorithms. These include methods such as those found in Weka \cite{abdalkareem2020machine}, oversampling techniques like SMOTE to increase the number of instances in the minority class \cite{chawla2002smote}, and under-sampling techniques like TOMEK \cite{dalla2021within}. Additionally, researchers can utilize a combination of both over- and under-sampling methods through the use of the SMOGN technique \cite{sharif2021deeporder}.
On the other hand, the currently available data sets in the context of CI need to be cleaned by removing unrelated data \cite{lee2019classifying} or noises \cite{byrne2020praxi}. Yaraghi et al.  \cite{yaraghi2022scalable}, employed a novel method for data preparation and created a dependency graph in which nodes refer to a source file and each edge shows a dependency relation from the source node to the destination to determine the code coverage of test cases.
% They proposed the method of building the data based on gathered data as a means of data preparation. This method involved creating a dependency graph to determine the code coverage of test cases and subsequently scoring the test cases based on this coverage.

% \par \textbf{Feature engineering:}
\par \subsubsection{Feature engineering}
It is important to normalize the feature values in a dataset due to the diversity in their distributions and range. This can be done by using statistical methods such as re-scaling the values to fit within a specific scale, like [0,1] for normalization \cite{grano2019branch}, standardizing by dividing the feature values by the standard deviation of all values \cite{parry2022evaluating}, or applying techniques such as log transformation to bring the values to the same magnitude and reducing the effect of extremely high or low values \cite{grano2018high}.
Scaling is commonly employed (17 out of 44 studies) in the context of CI due to the diversity of feature values and the availability of numerical features. Additionally, for converting text-based values to input that is understandable for ML methods, such as system logs \cite{byrne2020praxi} and the source codes \cite{lee2019classifying}, tagging and tokenization techniques are often used (9 out of 44). Word2vec \cite{lee2019classifying} and Columbus \cite{byrne2020praxi} methods are employed in the identified state-of-the-art studies for extracting tags from the texts. TF-IDF (term frequency-inverse document frequency) is a technique used to weigh the importance of words in a normalized way in texts \cite{abdalkareem2020machine}. It can help address the issue of imbalanced class distribution in text data by giving more weight to terms that appeared less frequently.

% \par \textbf{Learning algorithms:}
\par \subsubsection{Learning algorithms}
As shown in Table \ref{table:MLMethods}, all of the current state-of-the-art studies utilize supervised learning methods. Additionally, it should be noted that all papers published between 2020 and the time of writing this paper have employed supervised machine learning algorithms. This trend suggests that supervised methods are becoming more popular among researchers in recent years. The popularity of supervised learning can be attributed to the acceptable accuracy of the models and the availability of labelled data from automated tools in CI environments.

\par Also, Table \ref{table:MLMethods} shows that tree-based algorithms such as Decision Trees and Random Forests have been commonly employed as classification methods in state-of-the-art solutions. Tree-based algorithms have also been used in 19 studies from 44 selected studies. The reasons for the popularity of tree-based algorithms are threefold.
First, a CI environment continuously produces a huge volume of data, and ML models need to be retrained frequently \cite{yaraghi2022scalable}. Since training tree-based algorithms require low computational resources, it is suitable to employ them in practical CI environments \cite{al2020effect}.
Second, the performance of the tree-based algorithms is generally high in classifying unseen data \cite{finlay2014data}. For instance, for automating the TCP task in CI, Yaraghi et al. \cite{yaraghi2022scalable} compared their tree-based (XGBoost) model with BERT \cite{devlin2018bert}, the state-of-the-art language representation model. They found that their method achieved relatively similar results as BERT while the training cost of their model was lower than BERT.
Third, the tree-based algorithms can be well-interpreted, and thus they are understandable for humans \cite{witten2002data}.

\par According to Table \ref{table:MLMethods}, besides tree-based algorithms, Neural Network (NN) algorithms, including Long Short Term Memory (LSTM), Multi-Layer Perceptron (MLP), Recursive Neural Network (RNN), and Deep Neural Network (DNN) have also been commonly employed in six studies. One of the main reasons for the popularity of NN algorithms, despite their need for high computational resources for training, is their ability to automatically extract features from data sets and make accurate predictions \cite{porres2020automatic}.

% \par \textbf{Evaluation methods:}
\begin{table}
%\footnotesize
%\resizebox{\textwidth}{!}{%
\caption{Commonly employed performance measures, and their descriptions and formulas.}
\vspace{-2mm}
\begin{tabular}{{p{0.1\linewidth} p{0.5\linewidth} p{0.25\linewidth}}}
\hline
  \textbf{Measure} &
  \textbf{Description} & \textbf{Formula} \\ \hline
Precision & The percentage of the detected positive instances that were correct               & $\frac{TP}{TP+FP}$ \\[6pt]
Recall    & The proportion of positive instances that were correctly identified               & $\frac{TP}{TP+FN}$    \\[6pt]
F1-score  & The harmonic mean of recall and precision                                           & $\frac{2 \times Precision \times Recall}{Precision+ Recall}$  \\[6pt]
Accuracy  & Percentage of correctly classified instances                                        & $\frac{TP+TN}{TP+FP+FN+TN}$  \\[6pt]

\hline
 \multicolumn{3}{l}
{\makecell[l]{\textbf{Notes:} TP=True Positives, TN=True Negative,FP=False Positives,\\ FN=False Negatives.}}
\end{tabular}
%}
\label{table:PerfMeas}
\vspace{-7mm}
\end{table}
\par \subsubsection{Evaluation methods}
% Regarding evaluation methods, K-fold methods have been used in 19 out of 44 selected studies. This technique is based on dividing data into K equal segments (folds), and then in each of the K evaluation rounds, ML models are trained based on the K-1 segments of the data and evaluated on the remaining unseen segment of the data set.
The K-fold method is widely used for evaluating ML models. It involves dividing the data into K equal segments (folds), and then in each of the K evaluation rounds, the model is trained on K-1 segments of the data and evaluated on the remaining unseen segment. According to the selected studies, 19 out of 44 have used this technique.
In general, evaluation methods are classified into random and sorted sampling. Under random sampling, each row of data has the same possibility to be selected as training, testing, and holdout data set, while in sorted sampling the testing set must be newer than the training data set. In sorted K-fold methods, the unseen testing data folds must be selected from the earlier folds in comparison with the training data set.
\par F1-score, precision, and recall are commonly used evaluation measures in literature, with 16, 20, and 23 studies utilizing them respectively. They are the most frequently used metrics in the field. The description and formula of these measures are presented in Table \ref{table:PerfMeas}. Also, AUC (``Area under the ROC Curve'') has also been used, in which ROC is a graph on the plot of True Positive Rate (TPR or Recall) as y-axis and False Positive Rate ($\frac{FP}{FP+TN}$) as the x-axis.

% \begin{equation}
% \small
% Precision = \frac{TP}{TP+FP}
% \label{Eq:Precision}
% \end{equation}

% \begin{equation}
% \small
% Recall = \frac{TP}{TP+FN}
% \label{Eq:Recall}
% \end{equation}

% \begin{equation}
% \small
% F1-Score = \frac{2 \times Precision \times Recall}{Precision+ Recall}
% \label{Eq:F1}
% \end{equation}

% \begin{equation}
% \small
% FPR = \frac{FP}{FP+TN}
% \label{Eq:FPR}
% \end{equation}

% \begin{equation}
% \small
% Accuracy = \frac{TP+TN}{TP+FP+FN+TN}
% \label{Eq:Accuracy}
% \end{equation}

The evaluation measure of efficiency (cost) has received significantly less attention compared to measures of effectiveness. Specifically, only Yaraghi et al.\cite{yaraghi2022scalable} reported cost-related evaluation metrics, which know as Cost-cognizant Average Percentage of Faults Detected ($APFD_{c}$). Malishevsky  et al.~\cite{malishevsky2006cost} presented another version of $APFD_{c}$ that combines the cost of detecting and the severity of faults as presented in Eq.~\eqref{Eq:APFDC}. In this metric $n$ is the number of test cases in test suite $T$ with costs $t_{i}$ and $f_{i}$ is the severity of faults. $TF_{i}$ is the first test case that reveals the fault $i$.

\begin{equation}
\small
APFD_{c} = \frac{\sum_{i=1}^m(f_{i} \times (\sum_{j=TF_{i}}^n t_{j}-\frac{1}{2}t_{TF_{i}}))}{\sum_{i=1}^n t_{i} \times \sum_{i=1}^m f_{i}}
\label{Eq:APFDC}
\end{equation}

\section{Conclusion and Future Opportunities}
In this study, we distilled the CI phases automated by ML and the respective state-of-the-art ML solutions in this emerging area.
% According to the results of this study, we found that the state-of-the-art methods utilized
We found data from multiple open-source projects were commonly considered for developing ML solutions. Additionally, preprocessing techniques such as filtering and balancing input data and scaling feature values were used to improve the model performance. The use of tree-based algorithms was common among the selected studies, as they are capable of extracting patterns from large-scale CI data with low computational overhead. Furthermore, we highlighted a variety of evaluation methods and measures employed, to facilitate the comparison between future work and previous studies.
Despite the promising results of the current ML solutions, we identify three key challenges that may hinder the real-world adoption of these solutions and suggest potential directions to address these challenges.
% Besides the aforementioned points,
% the characteristics of the employed techniques in the training and evaluation phases of the ML methods, and the input data are diverse in the state-of-the-art methods. Despite these diversities, 
% we discuss the gaps and weak points of the reviewed studies.
% a few gaps and weak points can be observed in the reviewed studies.
% Considering these points in future studies will improve the applicability and reliability of the presented methods in real-world problems.

% \textbf{Realistic evaluation:}
\textit{Realistic evaluation:}
Despite the known limitations of random-based evaluation methods, these techniques have been utilized in a notable number of studies in the field. Specifically, five state-of-the-art studies and 18 studies out of the 44 reviewed in the literature have employed these methods. Random-based evaluation methods have limitations and may not be reliable in real-world settings as they use future data for training the model, unlike the chronological data generation in CI environments, leading to a lack of robustness and generalizability of results. Thus, ML models must be trained only on current data and evaluated on new and unseen data to ensure a realistic evaluation of the model performance when deployed in real-world environments \cite{elsner2021empirically}.
% Additionally, the risk of over-fitting for the trained ML model is reported as a disadvantage of random sampling K-fold  \cite{grano2019branch}.

% \textbf{Cost benchmark:}
\textit{Cost benchmark:}
In addition to reporting performance and effectiveness measures, there is a need for reporting cost measures. These cost measures include the time to train and the time to perform prediction/classification.
% These measures provide a better perspective for researchers and software engineers for future studies and decision-making.
Cost-based metrics only have been used in \cite{yaraghi2022scalable}, from all the selected state-of-the-art studies. However, they did not report the training and prediction cost/time of their presented models. Therefore, a fully cost-benefit analysis can be taken by combining performance and cost measures. Such analysis can provide more details about potential trade-offs when adopting the presented models in industrial environments.
\par\textit{Under-explored CI areas:}
According to the reviewed studies, the application of ML methods in the CI testing environment is limited to nine areas, as explained in Figure~\ref{Fig:CITasks}. 
A recent review study on predictive models in software engineering, Yang et al. \cite{yang2022predictive}, highlights the potential for applying ML-based methods in various software engineering testing areas. However, upon further examination, it is apparent that there are several under-explored software engineering testing areas that have yet to be fully explored in the context of CI literature. A number of them are including crash prediction, erroneous behaviour detection, bug severity classification, test report classification and assessment, fault injection, bug report management tasks including bug report assignment, bug report prediction and classification, software quality assessment and reliability, vulnerability and malware detection, code smell detection, traceability detection methods for bugs, and code clone detection.
It is still unclear to what extent these areas and their respective ML solutions can be adapted to CI. Thus, it is suggested that researchers and practitioners should also explore these missing areas in addition to the current ones to maximize the utilization of ML for CI automation.
% Based on reviewing \cite{yang2022predictive} and the selected papers in our study 

\section*{Acknowledgement}
We acknowledge the contribution of Dr Zohaib Md. Jan and Roshan Namal Rajapakse during the first phase of collecting and analysing data in this study.
% We would like to thank Dr. Zohaib Md. Jan for his valuable help in gathering data for this study.

\bibliographystyle{ieeetr}

\bibliography{ms}

% \bibitem{b1} G. Eason, B. Noble, and I. N. Sneddon, ``On certain integrals of Lipschitz-Hankel type involving products of Bessel functions,'' Phil. Trans. Roy. Soc. London, vol. A247, pp. 529--551, April 1955.
% \bibitem{b2} J. Clerk Maxwell, A Treatise on Electricity and Magnetism, 3rd ed., vol. 2. Oxford: Clarendon, 1892, pp.68--73.
% \bibitem{b3} I. S. Jacobs and C. P. Bean, ``Fine particles, thin films and exchange anisotropy,'' in Magnetism, vol. III, G. T. Rado and H. Suhl, Eds. New York: Academic, 1963, pp. 271--350.
% \bibitem{b4} K. Elissa, ``Title of paper if known,'' unpublished.
% \bibitem{b5} R. Nicole, ``Title of paper with only first word capitalized,'' J. Name Stand. Abbrev., in press.
% \bibitem{b6} Y. Yorozu, M. Hirano, K. Oka, and Y. Tagawa, ``Electron spectroscopy studies on magneto-optical media and plastic substrate interface,'' IEEE Transl. J. Magn. Japan, vol. 2, pp. 740--741, August 1987 [Digests 9th Annual Conf. Magnetics Japan, p. 301, 1982].
% \bibitem{b7} M. Young, The Technical Writer's Handbook. Mill Valley, CA: University Science, 1989.
% \end{thebibliography}

\end{document}